\documentstyle[aps,epsf,preprint]{revtex}

\begin{document}

\title{
Neutron and proton spectra from the decay of $\Lambda$
hypernuclei
}

\author{A. Ramos}

\address{Departament d'Estructura i Constituents de la Mat\`eria,
Universitat de Barcelona, 08028 Barcelona, Spain}

\author{M.J. Vicente-Vacas and E. Oset}

\address{
Departamento de F\'{\i}sica Te\'orica and IFIC, Centro Mixto
Universidad de Valencia-CSIC, 46100 Burjassot, Valencia, Spain
}

\maketitle

\begin{abstract}
We have determined the spectra of neutrons and protons following
the decay of $\Lambda$ hypernuclei through the one- and two-nucleon
induced mechanisms. The
momentum distributions of the primary nucleons are calculated and
a Monte Carlo simulation is used to account for
final state interactions. From the spectra we calculate the number
of neutrons
($N_n$) and protons ($N_p$) per $\Lambda$ decay and
show how the measurement of these quantities, particularly $N_p$, can
lead to a determination of $\Gamma_n / \Gamma_p$, the ratio of neutron
to proton induced $\Lambda$ decay.
We also show that the consideration of the two-nucleon
induced channel has a repercussion in the results, widening the band of
allowed values of $\Gamma_n / \Gamma_p$ with respect to
what is obtained neglecting this channel.
\end{abstract}

\newpage

\section{Introduction}

The apparent discrepancy between the ratio of the $\Lambda n \rightarrow
n n $ and $\Lambda p \rightarrow n p$ cross sections predicted by the one
pion exchange model (OPE) \cite{BRA,DUB,OSER} and experiments looking
at neutrons and  protons from the decay of $\Lambda$ hypernuclei
\cite{MON,BAR,EJI},
has stimulated much theoretical work, refining the OPE model with the
addition of the exchange of other mesons \cite{DUB,RAM1}, using
correlated two pion exchange \cite{RUS,MOT}, or considering quark
degrees
of freedom \cite{HED,OKA}. In spite of these efforts, no theoretical
approach leads, without ambiguities, to values of $\Gamma_n / \Gamma_p$
of the order of unity, as claimed by experiment, compared to values of the
order of 0.1 that one obtains in the OPE model.
Before proceeding further one should, however, notice that the
experimental errors are very large\cite{BAR,EJI}.
%and in some cases make the results
%compatible with values of $\Gamma_n / \Gamma_p$ as low as 0.1---0.2
%\cite{BAR,EJI}.

Some hopes for the understanding of the large fraction of neutrons observed
in the experiment were raised in Ref. \cite{WAN}, where the
two-nucleon (2N) induced $\Lambda$ decay
was studied. Indeed, in this latter mechanism
a near on-shell pion is produced in the $\Lambda N \pi$ vertex and
the pion is then absorbed mostly by a neutron-proton pair. Hence, one
has
the reaction $\Lambda n p \rightarrow n n p$ and two neutrons and one
proton are emitted in this process, while in the one-nucleon (1N)
induced $\Lambda$ decay processes ($\Lambda p \rightarrow n p$,
$\Lambda n \rightarrow n n$),
a $np$ or $nn$ pair is produced. Thus, even
if the $\Lambda n \rightarrow nn$ process was suppressed, as in the OPE
model, one could still have a large fraction of neutrons produced if
one had a sizable $\Lambda$ decay width through the 2N induced
channel.

The idea of Ref. \cite{WAN} was reanalyzed and the calculations
improved in Ref.
\cite{RAM2} with the surprising result that the consideration of the
2N induced channel, in connection with the measured number
of neutrons and protons, led to ratios of $\Gamma_n / \Gamma_p$ even
larger than those extracted before and in worse agreement with the OPE
results. These conclusions, however, were based on the assumption that
all the emitted particles were detected.

On the other hand, it was pointed out that, in case only two
particles from the $\Lambda N N \rightarrow N N N$ reaction were
detected, the analysis of the data with the consideration of the 2N
induced $\Lambda$ decay channel would lead to smaller values of
$\Gamma_n / \Gamma_p$\cite{GAL}.

These findings indicate that the values obtained for
$\Gamma_n / \Gamma_p$ are sensitive to the detection thresholds
for nucleons and,
hence, a precise determination of $\Gamma_n / \Gamma_p$ requires a
theoretical calculation of the nucleon spectra coming from the
different mechanisms. Furthermore, a precise comparison with the
experimental spectra
requires also to address the problem
of the final state interactions of the nucleons on their way out
of the nucleus.

The aim of the present paper is to evaluate the energy spectra for
protons and neutrons, coming from the different mechanisms, as 
functions of $\Gamma_n / \Gamma_p$, which will be treated as
an unknown. Hence, comparison of the experimental neutron and proton spectra
with the theoretical predictions, or even just the
number of neutrons and protons emitted per $\Lambda$ decay, would
allow to determine the ratio $\Gamma_n / \Gamma_p$.

\section{Determination of the initial nucleon spectra}

In order to determine the neutron and proton spectra we use the formalism of
Refs. \cite{RAM2,SAL} for the 1N and 2N
induced $\Lambda$
decay, which uses the local density approximation to evaluate the
width of finite $\Lambda$ hypernuclei starting from the self-energy of a
$\Lambda$ particle in infinite nuclear matter.

The mesonic $\Lambda$
decay can also be treated in this way \cite{SAL}, but the sensitivity to the
nuclear and pion wave functions makes the finite nucleus treatment more
accurate \cite{ITO,MOT2,NIE,STR}.
In any case, there is no need to consider the nucleons emitted in the
mesonic decay since their energy
is around 5 MeV, which lies below any of the ordinary
detection thresholds.
One may
still wonder about secondary nucleons coming from pionic
$\Lambda$ decay at some point in the
nucleus with a subsequent absorption of this pion on its way out of
the nucleus.
However, the pions produced in the pionic $\Lambda$ decay have an
energy of around 20---30 MeV and they are weakly absorbed in the
nucleus. Moreover, due to Pauli blocking, these pions will be mainly
produced at the surface and, consequently, they will be even less
absorbed. From the analogous $(\gamma, \pi)$
reaction in $^{12}$C we can see that less than $10 \%$ of the
25 MeV primary produced pions are later on
absorbed
(compare  $\sigma_{\rm abs}$
and $\sigma_{\rm dir\, abs}$ of Fig. 8 in Ref. \cite{RAF2}).
Since the mesonic width is
$\Gamma_m = 0.3 \Gamma_{\Lambda}$ for $^{12}_\Lambda$C, where
$\Gamma_{\Lambda}$ is the free $\Lambda$ width,
then the fraction of reabsorbed pions would be about
$3 \%$ of $\Gamma_\Lambda$ or, equivalently,
less than $2 \%$
of the total $\Lambda$ width in $^{12}_\Lambda$C\cite{BAR}. This is a
negligible  amount and would be
further reduced in heavier nuclei since $\Gamma_m$ decreases very fast
with the mass of the hypernucleus.

The preceding discussion allows us to consider only
the nucleons coming from the 1N and 2N
induced $\Lambda$ decay. The local density formalism of Refs.
\cite{RAM2,SAL} is particularly suited to treat the final state
interaction
of the nucleons since these are produced at a certain point in the
nucleus and with a certain momentum. Then we can follow these nucleons
by means of a Monte Carlo computer simulation which takes into account
quasielastic nucleon collisions, pion production,
etc. The accuracy of this procedure to deal with nucleon propagation in
the nucleus
was established in Ref. \cite{RAF},
by comparing some results with the corresponding full Quantum
Mechanical ones.

Recalling the results of \cite{RAM2,SAL} we write the decay width of a
hypernucleus as

\begin{equation}
\Gamma = \int d^3 k \tilde{\rho} ({\bf k}) \Gamma ({\bf k}) \ ,
\end{equation}
with
$$
\Gamma ({\bf k}) = \int d^3 r |\psi_{\Lambda} ({\bf r})|^2
\Gamma ({\bf k},\rho ({\bf r})) \ ,
$$
where $\psi_{\Lambda} ({\bf r})$ is the $\Lambda$ wave function
in the nucleus and $\rho ({\bf r})$ the nuclear density.
The former equations show that $\Gamma ({\bf k})$ is evaluated by
means of the local density approximation and $\Gamma$ is then
obtained by weighing
$\Gamma {\bf (k)}$
with the momentum distribution of the $\Lambda $ in nucleus,
$\tilde{\rho} ({\bf k})$.
The weighing over ${\bf k}$ is more important in the evaluation of the
mesonic decay width
for heavy nuclei but has little relevance in the nonmesonic decay width
which we study here, since $\Gamma {\bf (k)}$ is rather
smoothly dependent on ${\bf k}$ for this channel. However,
keeping the ${\bf k}$  dependence becomes again relevant
for the distribution of momenta of
the emitted nucleons.

The nuclear matter width, $\Gamma ({\bf k},\rho)$, is
evaluated from the
$\Lambda$ self-energy via the equation
\begin{equation}
\Gamma = - 2 Im\, \Sigma
\end{equation}
where $\Sigma$ accounts for the diagrams of Figs. 1 and 2.
As shown in Refs. \cite{RAM2,SAL} the resulting width is
\begin{eqnarray}
\Gamma({\bf k},\rho) & = &- 6 (G \mu^2)^2 \int \frac{d^3 q}{(2 \pi)^3}
[ 1 - n ({\bf k} - {\bf q})]
\theta (k^0 - E ({\bf k} - {\bf q}) - V_N) \nonumber \\
& \times & Im \, \alpha  (q) |_{q^0 = k^0 - E ({\bf k} - {\bf q}) -
V_N}
\end{eqnarray}
with
\begin{eqnarray}
\alpha (q) &=&  \left( S^2 + \left(\frac{P}{\mu}\right)^2 {\bf q}^2
\right) F^2 (q) D_0 (q) \nonumber \\
  &+& \frac{\tilde{S}^2 (q) \overline{\Pi}^* (q)}{
1 - V_L (q) \overline{\Pi}^* (q)}
+ \frac{\tilde{P}\,^2_L (q) \overline{\Pi}^*  (q)}{1 - V_L (q)
\overline{\Pi}^* (q)}
 + 2 \frac{\tilde{P}^2_T (q) \overline{\Pi}^* (q)}{1 - V_T (q)
\overline{\Pi}^* (q)}  \ ,
\end{eqnarray}
where $\alpha(q)$ is related to the dressed pion propagator
and accounts for the effect of short range $NN$
and $\Lambda N$ correlations.
The explicit expressions for $\tilde{S}, \tilde{P}_L, \tilde{P}_T$
are defined in Eqs. (20), (23), and (24) of Ref. \cite{SAL} (denoted
there
by $C' , B', A'$ respectively).  In Eqs. (3) and (4), the quantities
$E({\bf p})$ and $V_N$
stand for the nucleon energy, $\sqrt{{\bf p}^2 + M^2}$, and potential
energy, respectively,
$F (q)$ is the $\pi NN$ form factor, $n ({\bf p})$ the nucleon
occupation number of a noninteracting Fermi system of density $\rho$,
$D_0 (q)$ the free pion
propagator, and $V_L$ ($V_T$) the longitudinal (transverse)
part of the spin-isospin $ph$ interaction.
The function
$\overline{\Pi}^*$, related to the pion self-energy through
\begin{equation}
\Pi^* (q) = \frac{f^2}{\mu^2} {\bf q}^2 F^2 (q) \overline{\Pi}^* (q)
\ ,
\end{equation}
is given by
\begin{equation}
\overline{\Pi}^*  = \overline{\Pi}^*_{\rm 1p1h} +
\overline{\Pi}^*_{\Delta {\rm h}} +
\overline{\Pi}^*_{\rm 2p 2h} \, ,
\end{equation}
and
accounts for particle-hole (1p1h), delta-hole ($\Delta$h)
and two particle-two hole (2p2h) excitation. Furthermore,
$\overline{\Pi}^*_{
\rm 1p 1h} = U_N$ and $\overline{\Pi}^*_{\Delta {\rm h}} =
U_\Delta$ are the
ordinary Lindhard functions for 1p1h and $\Delta$h excitation
\cite{WAL},
with the normalization of the appendix of Ref. \cite{OSER}. On the other
hand, $\overline{\Pi}^*_{\rm 2p 2h}$ is constructed in Ref.
\cite{RAM2} from data of p-wave
pion absorption in pionic atoms, extrapolated for pions off shell by means of
the phase space for real 2p2h excitation. In Eq. (3) we have
$Im \, \alpha (q)$ which, as one can see from Eq. (4), contains the
free pion propagator and other terms which renormalize the
pion in the medium. As noted in Ref. \cite{RAM2}, the sum of the
longitudinal terms from $Im \, \alpha (q)$ in Eq. (4) leads to a peak
around the position of the renormalized pion in the medium, with a
width given by
\begin{equation}
\Gamma_\pi (q) = - \frac{1}{\tilde{\omega} (q)} Im\, \Pi
(\tilde{\omega}(q), q) \ ,
\end{equation}
where $\Pi$ is the pion proper self-energy and $\tilde{\omega} (q)$
the renormalized pion energy in the medium.
The proper self-energy $\Pi$ is related to $\overline{\Pi}^*$ by means
of
\begin{equation}
\Pi (q^0, q) = \frac{\Pi^* (q^0, q)}{1 - \frac{f^2}{\mu^2}
g' (q) \overline{\Pi}^* (q^0, q)}
\end{equation}
with $g'$ the Landau-Migdal parameter (smoothly $q$ dependent).
For the pions emitted in the $\Lambda$ decay, $Im\, \Pi$ in Eq. (8) is
actually $Im\, \Pi_{\rm 2p 2h}$, since there is no strength from $Im\,
\Pi_{\rm 1p 1h}$
at the pion pole and furthermore $Im\, U_{\Delta}$ is practically zero
there.
One must then be cautious interpreting the strength  coming from
$Im\, \overline{\Pi}^*_{\rm 2p 2h}$ as due to 2p2h excitation,
since part of it belongs to the
excitation of the renormalized pion, which contributes to the mesonic channel. 
In heavy nuclei, where the
pionic decay mode is practically forbidden by Pauli blocking,
this association is clear, but in light and medium nuclei, where there is
still a certain fraction of mesonic decay, one must do the separation
of the mesonic and 2p2h channels. On the other hand, the
1p1h channel offers no problems because it does not mix with the
pion pole term. Thus, the strength coming from Eq. (3),
obtained by substituting in
$Im\, \alpha (q)$ 
\begin{equation}
Im\, \frac{\overline{\Pi}^* (q) }{1 - V_{L,T} (q) \overline{\Pi}^*
(q)}
\rightarrow
 \frac{Im\, \overline{\Pi}^*_{\rm 1p 1h} (q)}{|1 - V_{L,T} (q)
\overline{\Pi}^*
(q) |^2}
\end{equation}
and omitting the $D_0(q)$ term, corresponds to 1p1h excitation.

In Ref. \cite{RAM2} the strength of the 2p2h excitation was
obtained by subtracting from
the whole width, calculated with the full $Im\, \alpha (q)$,
the contribution of the mesonic
and the 1p1h excitation channels. The mesonic channel was
calculated with the
zero width
approximation at the position of the renormalized  pion pole
and the 1p1h excitation channel was obtained with
the procedure indicated in Eq. (9).

In the present work we have adopted a more practical procedure which
leads to the same
results. The strength around the renormalized pion pole has been
omitted by cutting
$Im\, \alpha (q)$
between $\tilde{\omega} - \lambda\Gamma_\pi$ and
$\tilde{\omega} + \lambda\Gamma_\pi$, with $\lambda=0.8$, a
value that has been adjusted to reproduce
the same 2p2h width as that obtained in Ref. \cite{RAM2}.
This eliminates the contribution of the mesonic channel and leaves only
those of the 1p1h and the 2p2h excitation channels.
After this cut is done, the  part of $\Gamma$
in Eq. (3) proportional to $Im\, \overline{\Pi}^*_{\rm 1p1h}$, through
Eq. (9), and the analogous one proportional to $Im\,
\overline{\Pi}^*_{\rm 2p2h}$
are now associated to the
1N induced and 2N induced $\Lambda$ decay,
respectively.

The evaluation of the final nucleon momenta proceeds in two steps.
First, we determine the distribution of momenta after the $\Lambda$
decay (primary step). Next, we consider the final state
interactions of the nucleons via a Monte Carlo simulation, which will
be discussed in the next section.

In order to determine the primary nucleon momenta let us look at the
structure of the integrals involved in the evaluation of $\Gamma$
\begin{equation}
\Gamma_{i} = \int d^3 k \int d^3 r \int d^3 q \dots Im\,
\overline{\Pi}^*_i (q^0=k^0-E({\bf k}-{\bf q})-V_N, q) \ ,
\end{equation}
with the index $i$ standing for 1p1h or 2p2h, or alternatively,
1N and 2N induced mechanisms. On the other hand,
we have from Fig. 3a
\begin{equation}
Im\, \overline{\Pi}^*_{\rm 1p1h} (q^0, q)\, \, \propto \int d^3 p \,
n({\bf p})
[1 - n ({\bf p} + {\bf q}) ] \delta (q^0 + E({\bf p}\,)
- E({\bf p} + {\bf q})) \ ,
\end{equation}
which
can be further simplified eliminating the $\delta$ function.
Furthermore, from Fig. 3b we have
\begin{eqnarray}
Im\, \overline{\Pi}^*_{\rm 2p 2h} (q^0, q) & \propto & \int d^4 k' \,
Im\, U_N (\frac{q}{2}
 + k', \rho) Im\, U_N (\frac{q}{2} - k', \rho) \nonumber \\
&\times&\theta (\frac{q^0}{2} + k' \,^0) \theta (\frac{q^0}{2} -
k'\,^0) \ ,
\end{eqnarray}
which can be further simplified as shown in Ref. \cite{RAM2}.

%The idea is to assign, for each of the integration points involved in
%Eq. (10), a value of momentum for the emitted nucleons which will
%then
%be followed in their propagation through the nucleus. The evaluation
%of the nucleon momentum distribution is easy, since,
In the 1p1h
mechanism, a change of variables in Eq. (11) can be performed leaving
the momentum of the emitted nucleon, ${\bf p}+{\bf q}$,
as integration variable.
In the 2p2h case, each of the Lindhard functions
appearing in Eq. (12) involves an integration over an internal hole
momentum, ${\bf p}_{h_1}$ and
${\bf p}_{h_2}$, respectively. As for the previous case, a change of
integration variables in terms of the two emitted nucleon momenta
(${\bf q}/2+{\bf k}^\prime+{\bf p}_{h_1}$ and
 ${\bf q}/2-{\bf k}^\prime+{\bf p}_{h_2}$) can be made.
Then, one can perform the integrations using the Monte
Carlo method, since the integrands are smooth once the pion peak
is removed. Each configuration point generated by the Monte Carlo
technique
corresponds to a set of momenta for the outgoing primary
nucleons.
%At the same time we also get the momentum of the nucleon coming
%from the $\Lambda$ vertex in Fig. 1 and 2, which is given in both
%cases
%by ${\bf k} - {\bf q}$. This latter procedure is the one we follow
%in practice.

%Next step is to decide whether the nucleons are protons or neutrons.
In the evaluation of the 1p1h and 2p2h induced $\Lambda$ decay
widths we have followed Ref. \cite{RAM2} and the 1p1h induced
channel is evaluated using the OPE model. This  could give a poor
description of the $\Gamma_n / \Gamma_p$ ratio, but so far, the
different attempts to improve on this ratio discussed in the Introduction
share one feature in common, which is that the decay rate is
barely changed with respect to that obtained with the OPE model.
Hence, we keep the probability obtained by the OPE model for the 1p1h
channel fixed, and take the ratio $\Gamma_n / \Gamma_p$ as a variable.
For the same reason, we also keep the nucleon momentum distribution
provided by the OPE model. A pure phase space calculation gives a very
similar shape for the momentum distributions.

%This assumption is also justified by the following observation: we
%have performed a pure phase-space calculation in which the transition
%amplitude $\Lambda N\to NN$ has been left as a constant all along the
%integration domain. Although, evidently, this type of calculation
%does not provide the proper normalization, we have obtained a very
%similar shape for the nucleon spectra, which indicates that the
%momentum distribution of the emitted nucleons does not depend too
%much on the details of the transition mechanism.

\section{Monte Carlo simulation}

In this section we show how the charge selection
and the propagation of the nucleons is done with the Monte
Carlo simulation.
We follow closely the steps developed in the study of inclusive
pionic reactions \cite{SAL2}, in $(\gamma, \pi)$ reactions in nuclei
\cite{RAF2} and in $(\gamma, N), (\gamma, NN), (\gamma, N \pi)$
photonuclear reactions \cite{RAF}.

In the first place a random number is generated which decides
whether we have a 1N induced event
or a 2N induced one, according to their
respective probabilities. Next we determine the momenta of
the primary nucleons emitted in the decay process.
This is done by generating random configurations which are
weighted by their corresponding probability according to the model
described in the previous section.
%With the distributions of nucleon momenta obtained as described in
%the former
%section, we generate random configurations weighted by the
%probability
%obtained for the distributions.
This gives us the momenta of nucleon 1
and nucleon 2 in Fig. 4, for the 1N induced mechanism,
and nucleons
1, 2, 3 in Fig. 5 for the 2N induced mechanisms.

Next we determine
the charge of the particles.
In Figs. 4 and 5 we show diagrammatically the neutron and protons which would
come out from the 1p1h and 2p2h mechanisms.
In the case of the 1N
induced process we generate a random number which decides
whether we have $\Lambda n \rightarrow nn $ or $\Lambda p \rightarrow
n p$, according to the probability $\Gamma_n / \Gamma_p$, which
we keep as a free parameter in the theory. In the case of
$\Lambda n \rightarrow nn$, each neutron is given one of the momenta
corresponding to nucleon 1 and nucleon 2 in Fig. 4a.
% which are
%generated random according to the distribution probability.
In the
case of $\Lambda p \rightarrow np$ (Fig. 4b)
we associate also random, with equal probability, the $n$ and $p$ to
nucleons 1 and 2 and viceversa.
With all these random decissions one has now an event corresponding to a
pair of nucleons, $p n$ or $nn$, with some definite momenta.

If the event was a 2N induced one
(Fig. 5), then we decide by means of a random number whether one has
the mechanism of  Fig. 5a or the one of Fig. 5b, taking into account
that the one of Fig. 5b has a probability twice as large as the mechanism
of Fig. 5a. In the case of the mechanism of Fig. 5a one still has to
do a
further decission, which is whether to place $pn$ in numbers 2,3 or
viceversa. This
is also decided random giving the same weight to the two
possibilities (as would come out in a model of absorption dominated by
the $\Delta$ excitation in the $\pi N$ vertex \cite{WEI}).
However, in this case this last step has no consequences since the
distribution
of nucleons 2 and 3 is symmetrical from Eq. (12).
%It just helps to
%make the distribution of events more homogeneous from the beginning.

With the former steps we have selected a configuration for a primary
event. One of the variables in the Monte Carlo integration is ${\bf r}$,
the vector position in the nucleus where the $\Lambda$ decay takes
place. Hence, each event in the Monte Carlo
integration determines the point at which the primary nucleons are
produced and this allows us to follow the fate of these nucleons on
their way out of the nucleus. This is done by allowing the nucleons
to undergo collisions with other nucleons of the nucleus according
to $NN$ cross sections, modified by Pauli blocking and polarization phenomena.
The method is detailed in Ref. \cite{RAF}, where a useful parametrization
of the cross sections borrowed from Ref. \cite{CUG} is also shown.

In the Monte Carlo simulation the nucleons emitted in the
decay are allowed to move through the nucleus under the influence of
a local potential given by the Thomas Fermi model,
$V_N (r) = - k_F (r)^2 /2M$. As the nucleons move out of the nucleus
they collide with the nucleons of the local Fermi sea.
In the collisions the nucleons change energy, direction and,
eventually, charge since the differencial cross sections used for $pn$
collisions allow configurations in which a fast proton colliding with
a neutron of the Fermi sea gives rise to a fast neutron and a slowly
moving proton. In each
collision, a nucleon from the Fermi sea is excited above the
local Fermi momentum and the propagation of this secondary unbound
nucleon,
which has an energy larger than its mass, must also be followed.
Eventually, each primary nucleon may produce several nucleons that
leave the nucleus.
At the end, we know the energy and
direction of each one of the emitted protons and neutrons, be primary
or secondary nucleons.

%collisions the
%nucleons change
%energy and direction and eventually charge, since the differential
%cross sections used for $pn$ collisions allow configurations in which
%a fast proton colliding with a neutron of the Fermi sea gives rise to a
%fast neutron and a slower proton.
%In each collision a nucleon from the Fermi sea is excited above it
%and manages to scape the nucleus (we use the $N$ potential energy of
%the Thomas Fermi model,  $V_N (r) = - k_F (r)^2 /2M$, such that a
%nucleon excited above the Fermi sea has total positive energy and
%hence can leave the nucleus). In this way, if the
%primary nucleons undergo collisions, several nucleons leave the
%nucleus at the end. We follow each of them and at the
%end we know the energy and direction of each one of the  emitted
%neutrons and protons, be primary or secondary nucleons.

\section{Results and discussion}

In Fig. 6 we show the spectrum of neutrons and protons coming
from the 1N and 2N induced mechanisms in the
decay of $^{12}_{\Lambda}$C, assuming a value $\Gamma_n/\Gamma_p=1$.
Hence, as can be inferred from Fig. 4
for the 1N induced mechanisms,
one expects three times more neutrons
than protons at each energy 
if one neglects the effect of final state
interaction and charge exchange. By comparing the dotted line
(neutrons) with the dashed line (protons), we can see
that this
is approximately the case, except at low energies where mostly secondary
nucleons show up. The kinetic energy peaks around 70 MeV and there is
a broad peak which reflects the Fermi motion of
 the nucleons and the $\Lambda$ momentum distribution.
 In the same figure,
the spectrum of neutrons (dash-dotted line) and protons (solid
line) coming from the 2N
induced mechanism is also shown. In this case,
the distribution
is rather flat, since three particles are involved in the initial and
final state and both Fermi motion and the final
phase space collaborate in producing the broadening of the spectrum.
At large
energies we find that there are about four times more neutrons than
protons. This indicates that the nucleons appearing at this high
energy
region are mainly those generated from the absorption of the virtual
pion
(nucleons 2 and 3 in Figs. 5a and 5b). In fact, if only the two
nucleons coming from the absorption of the virtual pion contributed
to this part of the spectrum, we would have
found five times more neutrons than protons. However, although
the nucleons coming from de $\Lambda$ vertex are in general slow,
there is a tail at large energies which lowers the ratio.
%Actually, from the two nucleons coming
%from the absorption of the virtual pion we expect five times
%more neutrons than protons. However, the nucleons coming from the
%$\Lambda$ vertex are in general slow, but there is a tail at
%large energies which lowers that ratio.
On the other hand, we also observe in
Fig. 6 a peak with about the
same number of neutrons as protons at
low energies around 10 MeV.
These nucleons come mostly from the
$\Lambda$ decay vertex and from final state interaction effects.
We should note, however, that at energies around and below 20 MeV our
spectra are not realistic. The semiclassical Monte Carlo
procedure becomes progressively less reliable at low energies and
other phenomena like evaporation etc.,
not considered by us, would come into play. However, this is of minor
importance here since ordinary experimental detection thresholds are
higher than this energy.

The effect of final state interactions (FSI) can be seen by comparing
the solid lines with the dashed lines in Fig. 7. We observe that FSI
affect
mostly the nucleon distributions at energies below 40 MeV.
%, but the
%effects at higher energies are small.
Only a small fraction
of the neutrons and protons at large energies is
removed due to FSI, whereas, at energies
below 40 MeV, strength from both the degradation of
the primary nucleons and the emission of secondary nucleons
is collected.

As a complementary information we show in Fig. 8 the spectrum of protons
from the 1N (dotted line) and 2N (dashed line) induced mechanisms,
calculated with a value $\Gamma_n / \Gamma_p = 0.1$.
The solid line is the total proton spectrum.
The features of the spectra
are similar to those in Fig. 6 although
the number
of protons or neutrons per $\Lambda$ decay emitted in one case or
the other is obviously rather different. For this reason we discuss
below these magnitudes as functions of $\Gamma_n /\Gamma_p$.

In the first place, we show in Figs. 9 and 10 the ratio of the
number of emitted neutrons to that of emitted protons per
$\Lambda$ decay, $N_n/N_p$,
as a function of $\Gamma_n / \Gamma_p$, omitting FSI and
assuming, respectively, that all particles are observed or that a
threshold cut of 40 MeV is applied in the detection energy.
The idea behind these curves is to facilitate the determination of the
ratio $\Gamma_n/\Gamma_p$ from the measured $N_n/N_p$ ratio.

We observe that the resulting $N_n/N_p$ ratio increases with
the value of $\Gamma_n / \Gamma_p$, but the results depend on whether
we consider
only the 1N induced mechanism (dashed line) or we include
also the 2N induced one (solid line).
In Fig. 9 we observe that, for a value of $\Gamma_n/ \Gamma_p =  0.5$,
the ratio $N_n / N_p$ is the same whether one
considers
the 1N induced mechanism only or both mechanisms.
However,
given an experimental value of $N_n / N_p$, the corresponding
value of $\Gamma_n / \Gamma_p$ including the 2N induced
decay
is bigger than the one obtained considering only the 1N
induced
mechanism if $\Gamma_n / \Gamma_p  > 0.5$. The situation is
reversed if
$\Gamma_n / \Gamma_p < 0.5$. This is exactly the result obtained
analytically in
Ref. \cite{RAM2}. However, if a cut of 40 MeV in the detection energy
is applied,
as can be seen in Fig. 10, the point where the two curves cross
appears at larger values of $\Gamma_n / \Gamma_p$ (in the figure at
$\Gamma_n / \Gamma_p = 1.3$). As follows from the  discussions above
in connection with Fig. 6, a cut of 40 MeV would eliminate mostly
the peak corresponding to the nucleons emitted from the
$\Lambda$ vertex. Therefore, the results in Fig. 10 follow the
tendency indicated
in Ref. \cite{GAL}, where it is shown that, if the nucleon from the
$\Lambda$ vertex is not observed, the two lines in the figure
would cross at a value of $\Gamma_n / \Gamma_p = 2$.

In Fig. 11 we show results including FSI
and taking threshold energy cuts of 0,
30 and 40 MeV. The last two cases allow to
compare our results with the
measurements of Ref. \cite{BAR}. The numbers found there, corrected as
indicated in Ref. \cite{RAM2,FRA},
were
$N_n^{\rm TOT}=N_0 N_n =3400 \pm 1100 $, $N_p^{\rm TOT}=N_0 N_p =
1270 \pm 180$,  for $E_{\rm cut} = 30$ MeV and $N_n^{\rm TOT} = 2530 \pm
1050$,
$N_p^{\rm TOT} = 1112 \pm 130 $, for $E_{\rm cut}$ = 40  MeV,
where $N_0$ is the total number of decay events. From the results
shown in Fig. 11 we can deduce that the band of allowed values of
$N_n^{\rm TOT}/N_p^{\rm TOT}=N_n / N_p$ corresponds to values of
$\Gamma_n / \Gamma_p$ in the range
0.15---2.0 for $E_{\rm cut} = 30$ MeV and 0.0---1.65 for
$E_{\rm cut}= 40$ MeV.
It is important to note that these results are even
compatible with the OPE predictions. We should also point out that
the inclusion of the
2N induced channel enlarges the band of allowed values at
both ends,
with respect to the results which would be obtained omitting
this channel. One can also see that
the effect of the 2N induced channel becomes smaller for higher
threshold detection energies, as can be easily understood from the
fact that the average energy of
the nucleons in the 2N induced channel is smaller than in the 1N
induced one.

The information contained in Fig. 11 indicates that it is
rather
difficult to extract  $\Gamma_n / \Gamma_p$ from the ratio of
neutrons to protons $N_n/N_p$ unless
this ratio is determined with
high precission. The fact that the relative error of the ratio $N_n
/ N_p$ is the sum of the
relative errors in $N_n$ and $N_p$, together with the fact that
usually neutrons are measured with
little precission, makes the uncertainty of this magnitude very
large and leads to large errors in $\Gamma_n / \Gamma_p$.

It is clear from the former considerations that the separate number
of protons and neutrons per $\Lambda $ decay would provide more
information. The gain is twofold: on the one hand the
individual relative errors of $N_n$ and $N_p$ are
smaller than for their ratio. On the other hand,
one has two pieces of information which will provide two
independent bands of allowed values of $\Gamma_n / \Gamma_p$.
The intersection of the two bands will give the final allowed region.
This procedure should give rise to more precise
determinations of $\Gamma_n / \Gamma_p$ in  the future. For this
purpose we present our predictions in Figs. 12 and 13.

In Fig. 12, we show
the number of neutrons per $\Lambda $ decay event, $N_n$, as a
function of $\Gamma_n / \Gamma_p$. From top to bottom the results
correspond to detection energy cuts of
0, 30 and 40 MeV. The dashed lines correspond to considering only the
1N induced decay, while the solid line includes also the
2N induced one.
The number of protons per $\Lambda$ decay event, $N_p$,
is shown, as a function of $\Gamma_n /\Gamma_p$, in Fig. 13  with
the same meaning as in Fig. 12.

The potential of these two figures to
determine $\Gamma_n / \Gamma_p$ can be shown with the following
example: From Figs. 12 and 13 we see that for $\Gamma_n/\Gamma_p=0.5$
and $E_{\rm cut}=30$ MeV we obtain $N_n=1.28$ and $N_p=0.62$. Assume
now a $10\%$ error in the values of $N_n$ and $N_p$. From Fig. 12
one obtains the range
$\Gamma_n /\Gamma_p  =$ 0.175---1.0, while
Fig. 13 gives the range $\Gamma_n / \Gamma_p =$ 0.35---0.75. On the
other hand, if
only the ratio $N_n / N_p$ from Fig. 11  were used, the
range of values would be $\Gamma_n / \Gamma_p =$ 0.2---0.75. We can
therefore see that $N_p$ is more selective than $N_n$ in
order
to determine $\Gamma_n / \Gamma_p$
and also more selective than the ratio $N_n/N_p$.
However, in this particular example we can see that
the ratio $N_n /N_p$ is more selective than the number $N_n$
itself. The fact that $N_n$ increases as $\Gamma_n / \Gamma_p$
increases, while
$N_p$ decreases, makes
the ratio $N_n / N_p$ a steeper function of $\Gamma_n / \Gamma_p$
than any of the numbers $N_n$ or $N_p$ and helps in
getting smaller errors for $\Gamma_n / \Gamma_p$. However, one has the
handicap that one must sum the relative errors of $N_n$ and $N_p$.

With other values and other errors we could have different situations
in which the measurements of $N_n$ could give additional information
to the one provided by $N_p$, but the former example indicates that the
measurement of $N_p$ is probably the most crutial magnitude
in order to determine $\Gamma_n / \Gamma_p$.

Unfortunately the experiment of Ref. \cite{BAR} does
not
provide the number of decay events corresponding to the total
number of neutrons and protons measured and therefore, only the
ratio can be used in our analysis.
On the other hand, the work of Ref. \cite{EJI} contains a spectrum of
protons but
it is conditioned by several cuts, efficiencies and geometries of the
detectors, and does not allow the extraction of $N_p$ nor can it
be compared to the spectrum which we have calculated.
However, the determination of $N_p$ is one of the aims
of the collaboration in Ref.
\cite{HAS} in the near future. As hinted by our observation above,
the determination of $N_p$ alone can provide as much information as
the combined measurement of $N_n$ and $N_p$. As we have shown, precise
measurements of $N_p$ with different cuts, which can be done if
the spectrum of protons is also known, would provide reliable values
for $\Gamma_n / \Gamma_p$.

\section{Conclusions}

We have evaluated the spectrum of neutrons and  protons following
the decay of $\Lambda$ hypernuclei. For this purpose we calculated the
momentum  distribution of the nucleons coming from the one nucleon
induced and two nucleon induced $\Lambda$ decay. Final
state interaction of the nucleons was also considered using
a Monte Carlo computer simulation technique, successfully applied
to other physical processes. By integrating over the energy
spectrum we can also obtain the number of neutrons and protons
for any energy cut in the nucleon detectors. We have seen that the
measurement of $N_n$ and $N_p$, the number of neutrons and
protons per $\Lambda$ decay, can be used to determine the
ratio $\Gamma_n/\Gamma_n$ reliably.
We observed that the value of $N_p$ was more selective in
determining the value of
$\Gamma_n / \Gamma_p$ than $N_n$ or the ratio $N_n /N_p$,
and this should serve as a guideline for future experiments.

The two nucleon induced $\Lambda$ decay channel was found relevant in the
analysis. Even if the fraction of this decay channel is only
30$\%$ of the free $\Lambda$ width, or 20$\%$ of the total
$\Lambda$ width in the nucleus\cite{RAM2}, it has some repercussion in
the determination of $\Gamma_n / \Gamma_p$ and, as a consequence,
enlarges the error band for $\Gamma_n / \Gamma_p$, obtained
from given values of $N_n / N_p$, with respect to a determination omitting
this channel in the analysis. Even then, the ratio $\Gamma_n / \Gamma_p$
can be determined reliably provided one can measure
$N_p$ and $N_n$ (particularly $N_p$) with sufficient
precission.

The analysis done here, and the figures presented, will allow
a direct determination of $\Gamma_n / \Gamma_p$ from future
measurement of $N_n$ and $N_p$, which in view of the results obtained
here should be encouraged.

\vspace{3cm}

\acknowledgements

We would like to acknowledge useful discussions with A. Gal, H. Ejiri,
O. Hashimoto,
T. Kishimoto, J. Nieves and M. Oka. This work has been partly
supported by DGICYT
contracts AEN 93-1205, PB92-071 and by the European Union  contract
No. CHRX-CT 93-0323.

\begin{figure}
\vskip 15 cm
\caption{
$\Lambda$ self-energy diagrams for the one nucleon induced
channel from Refs. [13,15].
%\cite{RAM2,SAL}.
The dotted line cuts the
states which are placed on shell in the evaluation of $Im\, \Sigma$.
}
\end{figure}

\begin{figure}
       \setlength{\unitlength}{1mm}
       \begin{picture}(100,180)
       \put(25,0){\epsfxsize=12cm \epsfbox{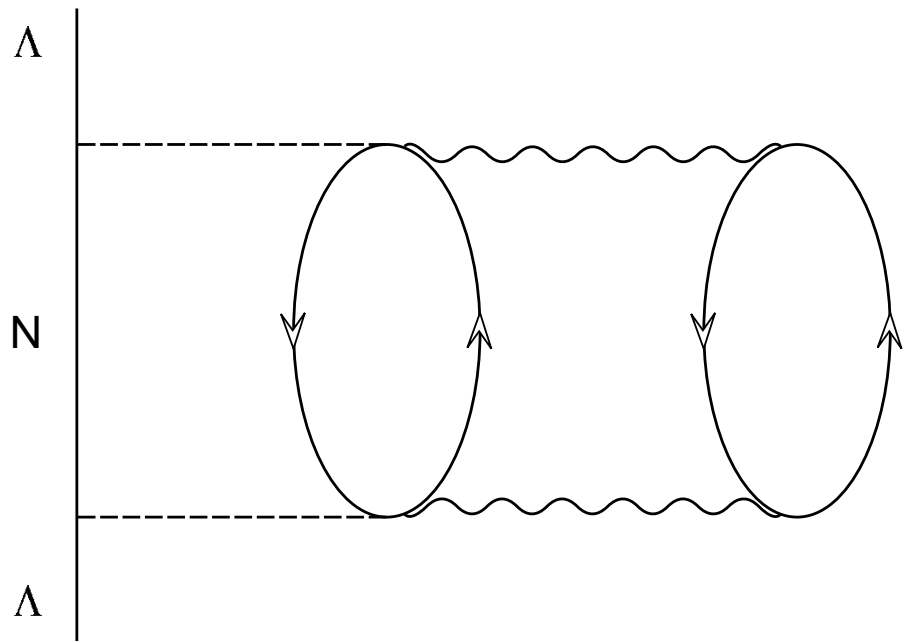}}
       \end{picture}
\caption{
$\Lambda$ self-energy diagram for the 2N induced $\Lambda$
decay mechanism from Ref. [13].
%\cite{RAM2}.
}
\end{figure}

\begin{figure}
       \setlength{\unitlength}{1mm}
       \begin{picture}(100,180)
       \put(25,0){\epsfxsize=12cm \epsfbox{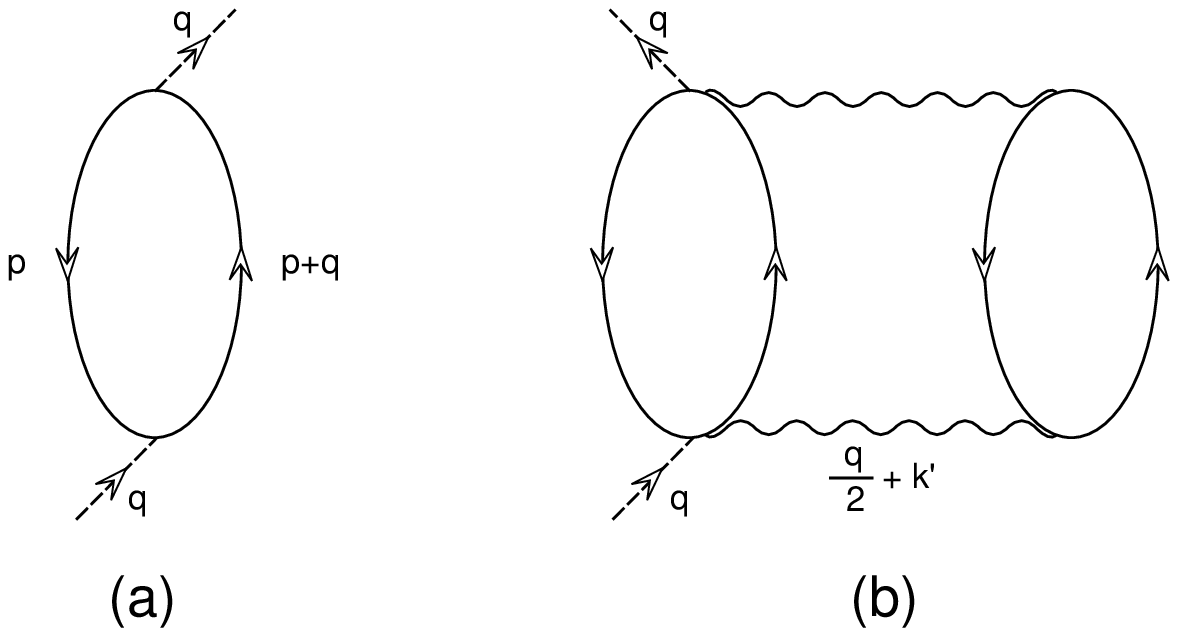}}
       \end{picture}
\caption{
 Diagrams employed in the evaluation of
 $Im\, \overline{\Pi}^*_{\rm 1p 1h}$ (a) and $Im\,
\overline{\Pi}^*_{\rm 2p 2h}$ (b).
}
\end{figure}

\begin{figure}
       \setlength{\unitlength}{1mm}
       \begin{picture}(100,180)
       \put(25,0){\epsfxsize=12cm \epsfbox{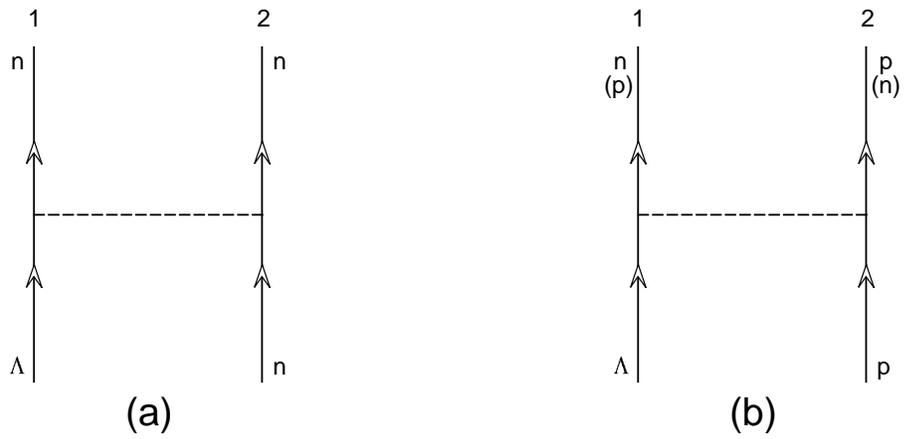}}
       \end{picture}
\caption{
 Feynman diagrams for the transition $\Lambda N
\rightarrow NN$:
neutron induced $\Lambda$ decay (a) and proton induced $\Lambda$
decay (b).
}
\end{figure}

\begin{figure}
       \setlength{\unitlength}{1mm}
       \begin{picture}(100,180)
       \put(25,0){\epsfxsize=12cm \epsfbox{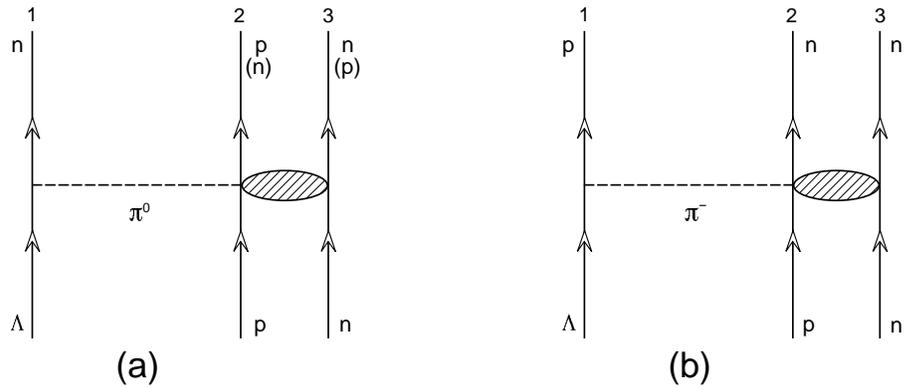}}
       \end{picture}
\caption{
 Feynman diagrams for the two nucleon induced $\Lambda$
decay through
virtual $\pi^0$ absorption (a) and virtual $\pi^-$
absorption (b).
}
\end{figure}

\begin{figure}
       \setlength{\unitlength}{1mm}
       \begin{picture}(100,180)
      \put(25,0){\epsfxsize=12cm \epsfbox{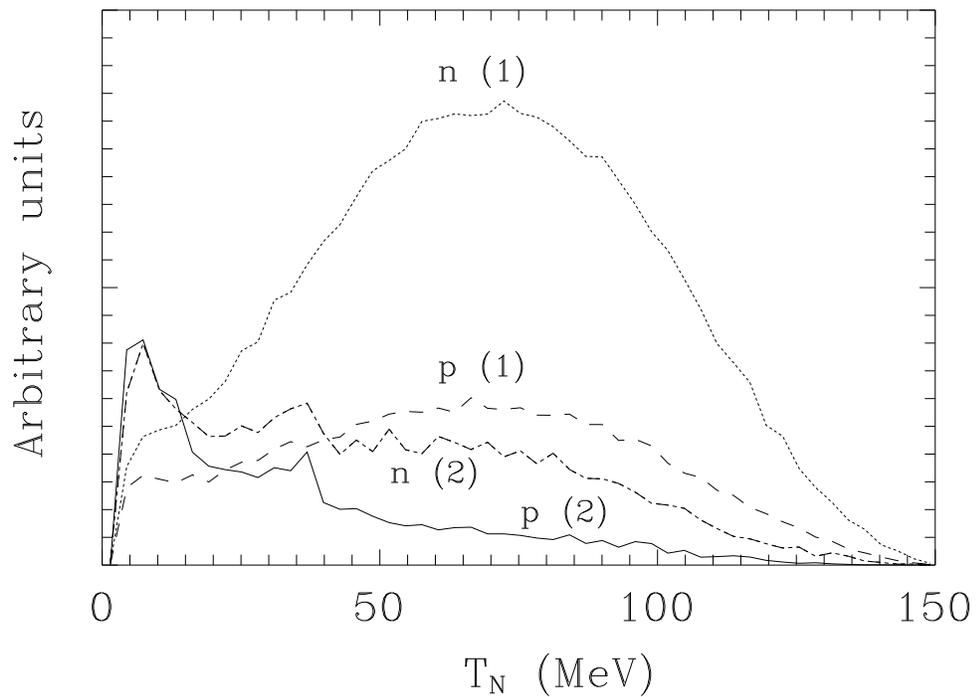}}
       \end{picture}
\caption{
 Spectra of neutrons and protons in the decay of
$^{12}_{\Lambda}$C.
Dashed line:
protons from the 1N induced mechanism. Dotted
line:
neutrons from the 1N induced mechanism. Solid line:
protons from the 2N induced mechanism. Dash-dotted line:
neutrons from the 2N induced mechanism. The results have
been obtained for a value $\Gamma_n/\Gamma_p = 1$.
}
\end{figure}

\begin{figure}
       \setlength{\unitlength}{1mm}
       \begin{picture}(100,180)
       \put(25,0){\epsfxsize=12cm \epsfbox{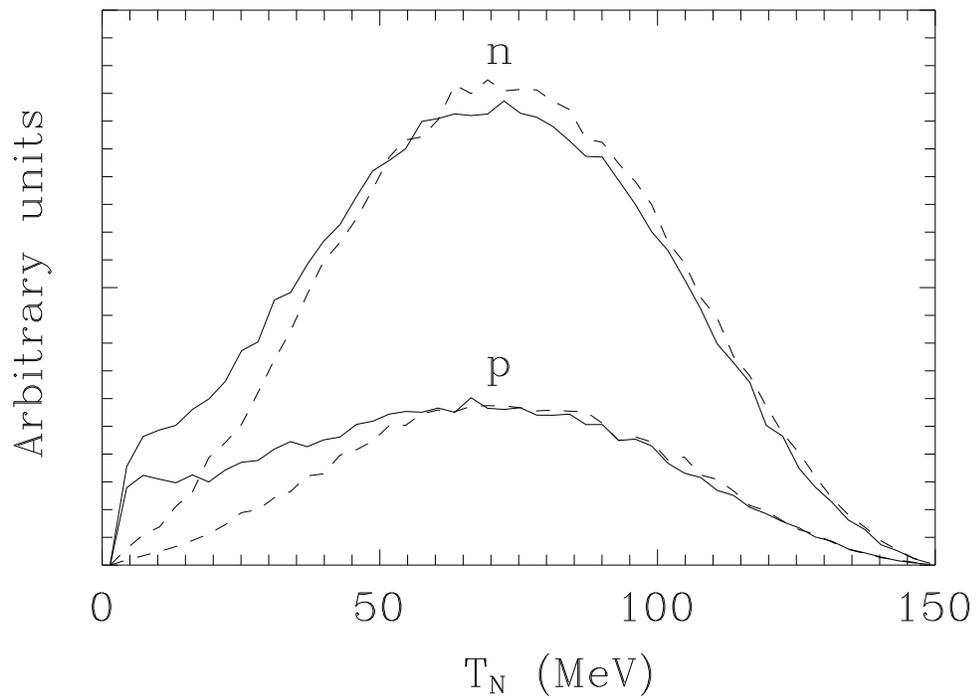}}
       \end{picture}
\caption{
Effect of the final state interactions in the spectrum
of nucleons emitted in the 1N induced decay.
Dashed line: results without FSI. Solid line: results
including FSI.
}
\end{figure}

\begin{figure}
       \setlength{\unitlength}{1mm}
       \begin{picture}(100,180)
       \put(25,0){\epsfxsize=12cm \epsfbox{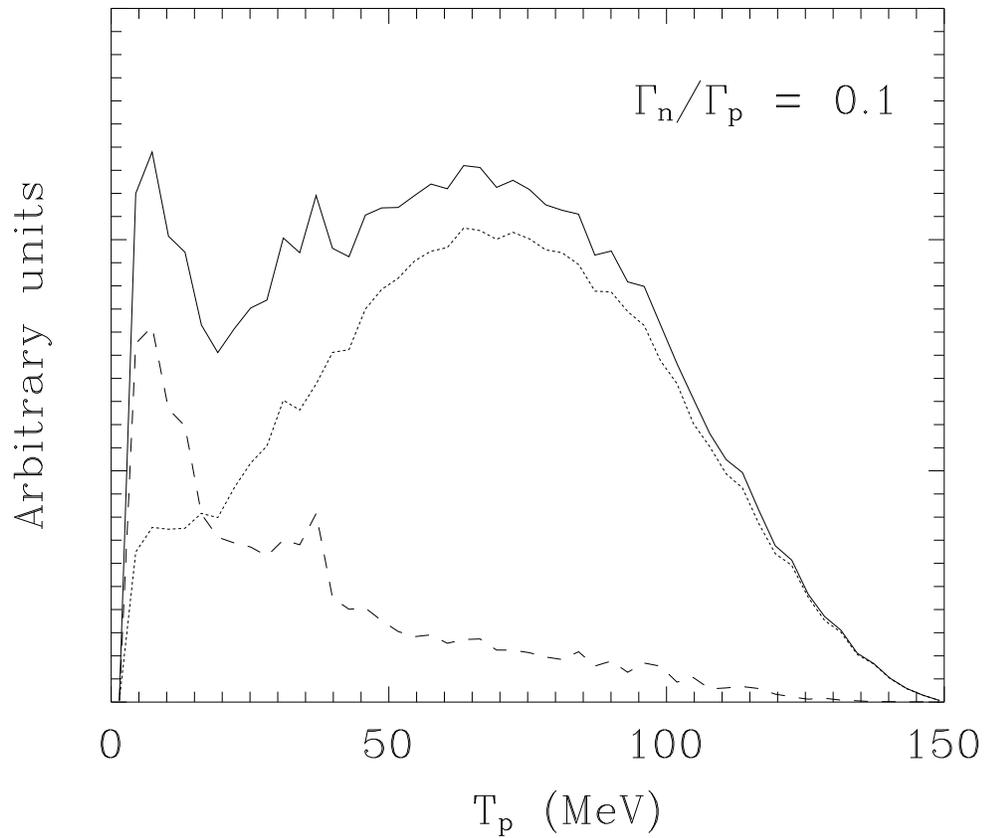}}
       \end{picture}
\caption{
 Proton spectrum obtained for a value $\Gamma_n/\Gamma_p =
0.1$. Dotted line:
1N induced mechanism. Dashed line: 2N induced mechanism.
Solid line: Total.
}
\end{figure}

\begin{figure}
       \setlength{\unitlength}{1mm}
       \begin{picture}(100,180)
       \put(25,0){\epsfxsize=12cm \epsfbox{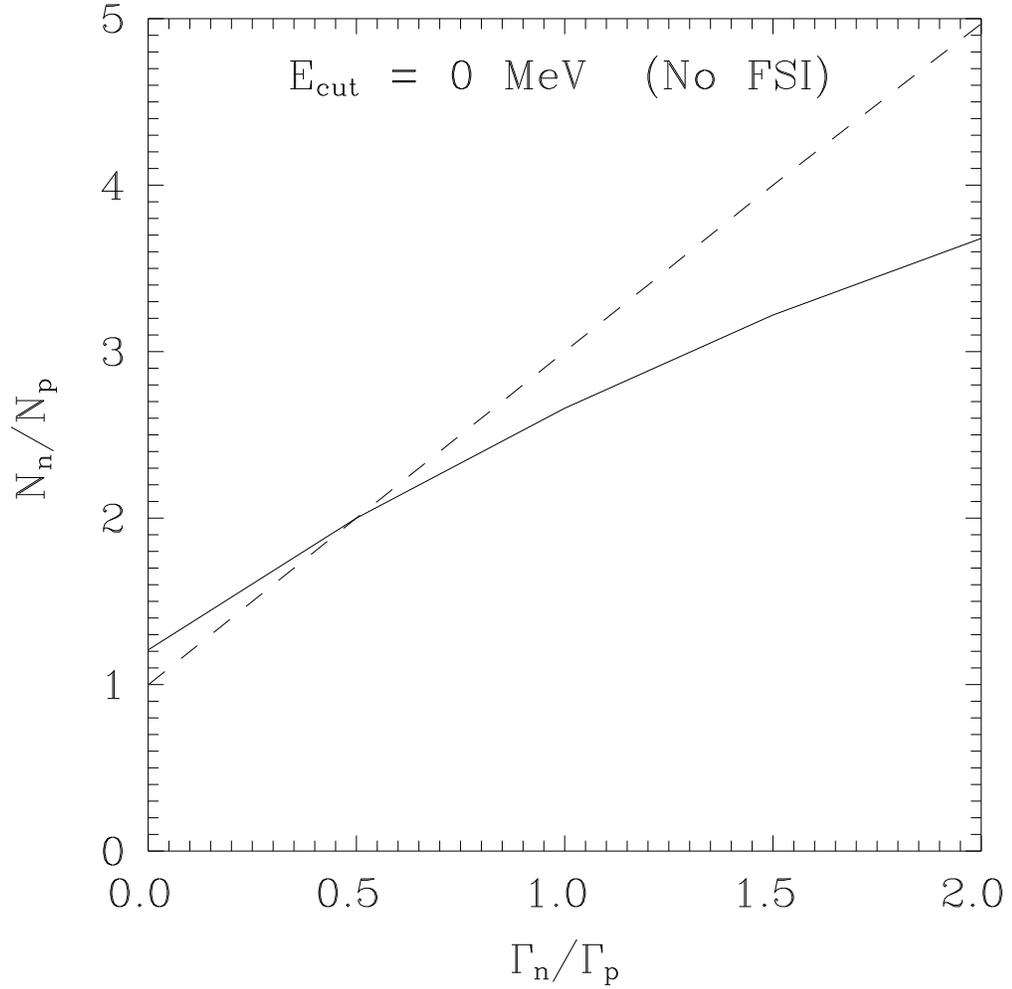}}
       \end{picture}
\caption{
 $N_n /N_p$ as a function of $\Gamma_n / \Gamma_p$ with no
cut in the detection energy and no final state interactions. Dashed
line: 1N
 induced mechanism. Solid line: 1N $+$ 2N induced mechanisms.
}
\end{figure}

\begin{figure}
       \setlength{\unitlength}{1mm}
       \begin{picture}(100,180)
       \put(25,0){\epsfxsize=12cm \epsfbox{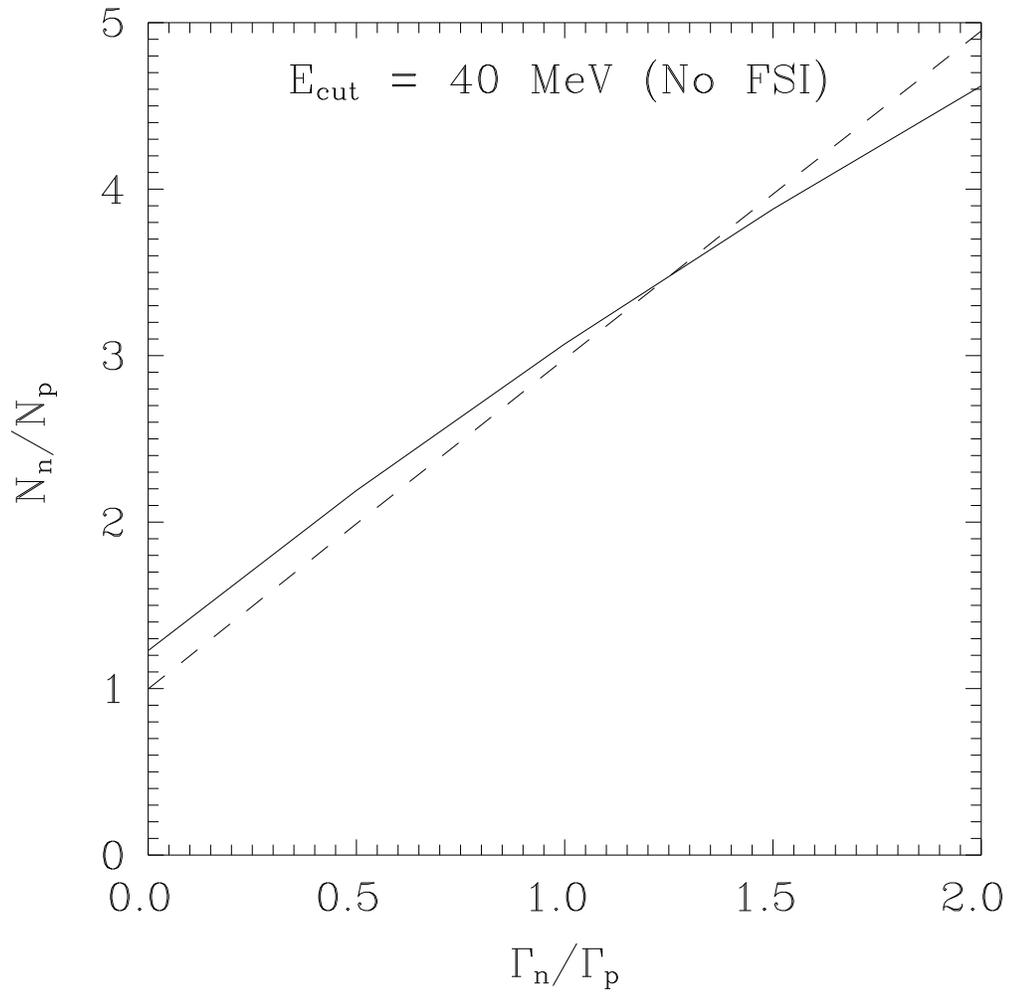}}
       \end{picture}
\caption{
 Same as Fig. 9, but with a detection threshold of 40 MeV and no
final state interactions.
}
\end{figure}

\begin{figure}
       \setlength{\unitlength}{1mm}
       \begin{picture}(100,180)
       \put(25,0){\epsfxsize=12cm \epsfbox{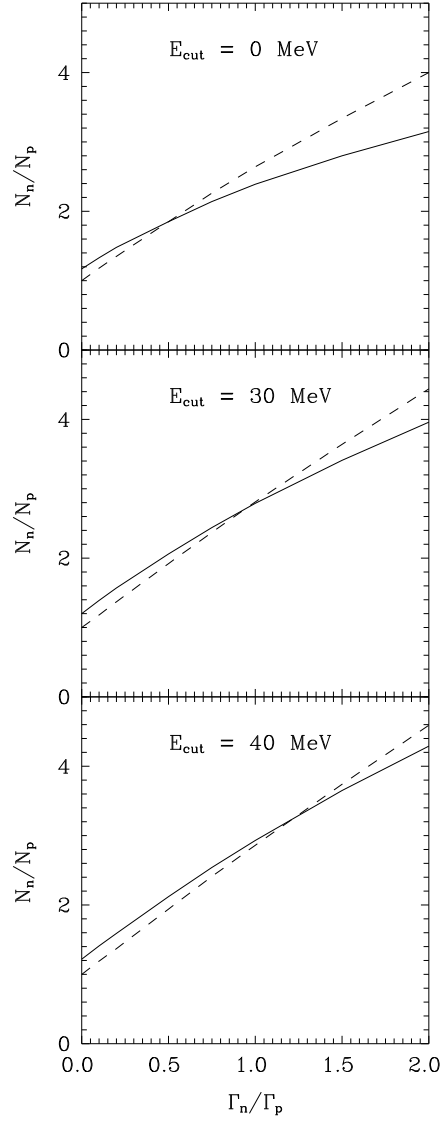}}
       \end{picture}
\caption{
 $N_n /N_p$ as a function of $\Gamma_n / \Gamma_p$ including final
state interaction effects and applying energy cuts of 0, 30 and 40
MeV. Dashed line: 1N induced mechanism. Solid line: 1N $+$ 2N
induced mechanisms.
}
\end{figure}

\begin{figure}
       \setlength{\unitlength}{1mm}
       \begin{picture}(100,180)
       \put(25,0){\epsfxsize=12cm \epsfbox{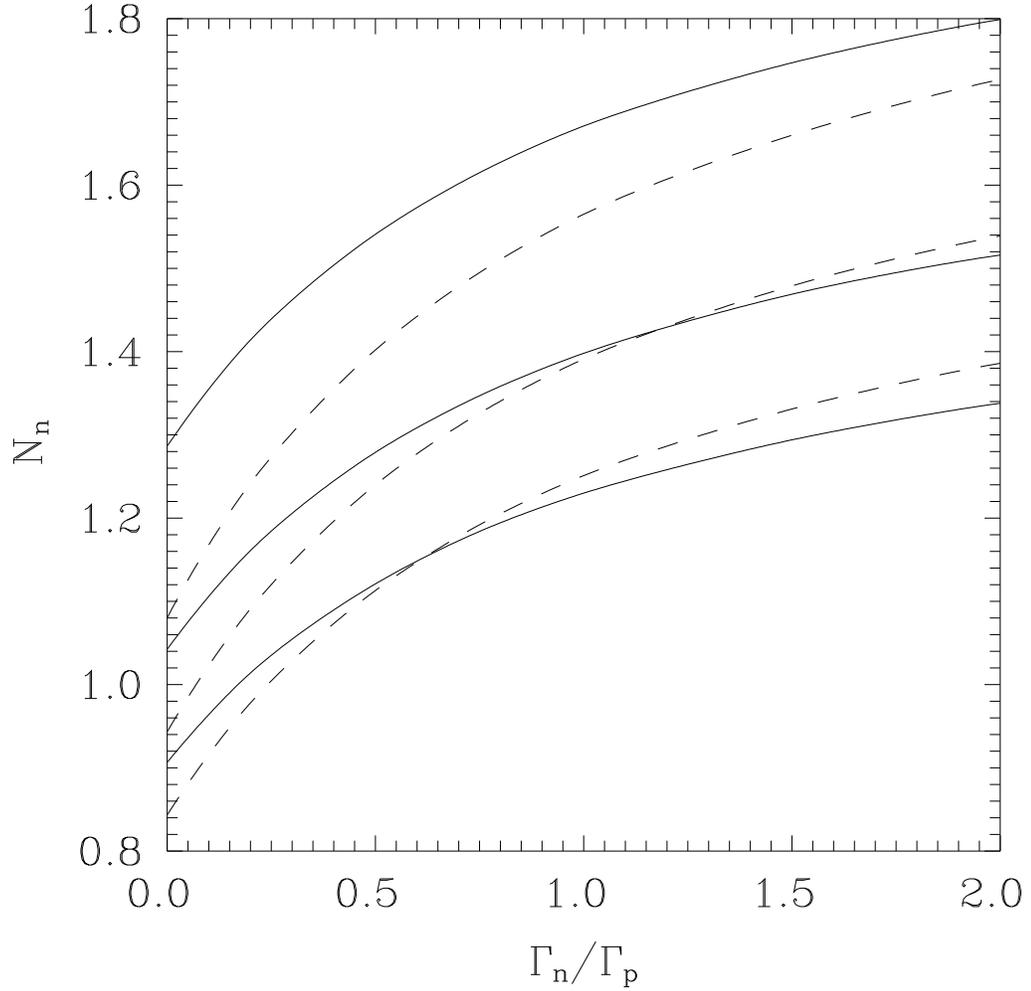}}
       \end{picture}
\caption{
 Number of neutrons per $\Lambda$ decay as a function of
$\Gamma_n / \Gamma_p$. Dashed lines: 1N induced mechanism. Solid
lines:
1N $+$ 2N induced mechanism. Final state interactions are considered
and, from top to bottom, the results include energy cuts of
0, 30 and 40 MeV, respectively.
}
\end{figure}

\begin{figure}
       \setlength{\unitlength}{1mm}
       \begin{picture}(100,180)
       \put(25,0){\epsfxsize=12cm \epsfbox{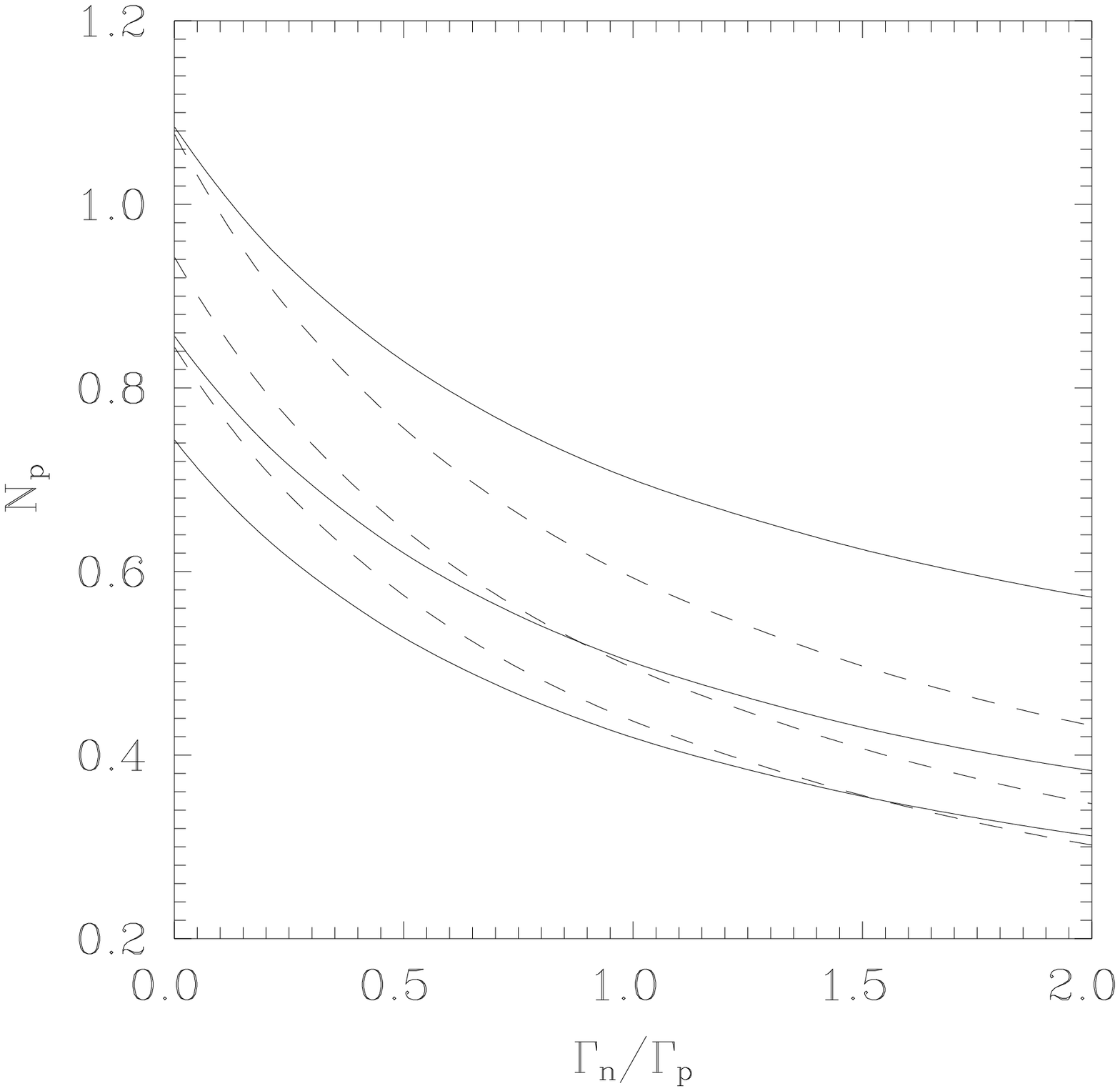}}
       \end{picture}
\caption{
 Same as Fig. 12 for the number of protons per $\Lambda$ decay.
}
\end{figure}


\begin{references}
\bibitem{BRA} K. Tacheuchi, H. Takaki and H. Bando, Prog. Theor. Phys.
{\bf 73}, 841 (1985).
\bibitem{DUB} J. Dubach, Nucl. Phys. {\bf A450}, 71c (1986).
\bibitem{OSER} E. Oset, P. Fern\'andez de C\'ordoba, L.L. Salcedo and
R. Brockmann, Phys. Rep. {\bf 89}, 79 (1990).
\bibitem{MON} A. Montwill et al. Nucl. Phys. {\bf A234}, 413 (1974).
\bibitem{BAR} J.J. Szymansky et al. Phys. Rev. C {\bf 43}, 849 (1991).
\bibitem{EJI} H. Noumi et al., Phys. Rev. C {\bf 52}, 2936 (1995).
\bibitem{RAM1} C. Bennhold, A. Parre\~{n}o and A. Ramos, Few
Body Systems Suppl. {\bf 9}, 475 (1995). A. Parre\~{n}o, A. Ramos and
C. Bennhold, Univ. of Barcelona preprint.
\bibitem{RUS} M. Shmatikov, Nucl. Phys. {\bf A580}, 538 (1994).
\bibitem{MOT} K. Itonaga, T. Ueda and T. Motoba, Nucl. Phys. {\bf
A585}, 331c (1995);
ibid, Proc. of the WEIN 95 Conference, H. Ejiri, T. Kishimoto and T. Sato
Eds., World Scientific 1995, p. 546.
\bibitem{HED} C. Y. Cheung, D.P. Heddle and L.S. Kisslinger, Phys. Rev.
C {\bf 27}, 335 (1983).
\bibitem{OKA} T. Inoue, S. Takeuchi and M. Oka, Nucl. Phys. {\bf
A597}, 563 (1996).
\bibitem{WAN} W. Alberico, A. de Pace, M. Ericson and A. Molinari,
Phys. Lett. {\bf B256}, 134 (1991).
\bibitem{RAM2} A. Ramos, E. Oset and L.L. Salcedo, Phys. Rev. C {\bf
50}, 2314 (1994).
\bibitem{GAL} A. Gal, in Proc. of the WEIN 95 Conference, H. Ejiri, T. Kishimoto
and T. Sato Eds., World Scientific 1995, p. 573.
\bibitem{SAL} E. Oset and L.L. Salcedo, Nuc. Phys. {\bf A443}, 704
(1985).
\bibitem{ITO} K. Itonaga, T. Motoba, and H. Bando, Z. Phys. {\bf
A330}, 683 (1988).
\bibitem{MOT2} T. Motoba, Nucl. Phys. {\bf A527}, 485c (1991);
Few Body Systems Suppl. {\bf 5}, 386 (1992); Nucl. Phys. {\bf A547},
115c (1992).
\bibitem{NIE} J. Nieves and E. Oset, Phys. Rev. C {\bf 47}, 1478
(1993).
\bibitem{STR} U. Straub, J. Nieves, A. Faessler and E. Oset, Nucl. Phys.
{\bf A556}, 531 (1993) 686.
\bibitem{RAF2} R.C. Carrasco, E. Oset and L.L. Salcedo, Nuc. Phys.
{\bf A541}, 585 (1992).
\bibitem{RAF} R.C. Carrasco, M.J. Vicente-Vacas and E. Oset, Nucl.
Phys. {\bf A570}, 701 (1994).
\bibitem{WAL} A.L. Fetter and J.D. Walecka, Quantum Theory of
Many Particle Systems (McGraw-Hill, N.Y. 1971).
\bibitem{SAL2} L.L. Salcedo, E. Oset, M.J. Vicente Vacas and C.
Garcia-Recio, Nucl. Phys. {\bf A484}, 557 (1988).
\bibitem{WEI} E. Oset and W. Weise, Nuc. Phys. {\bf A319}, 477 (1979).
\bibitem{CUG} J. Cugnon, private communication.
\bibitem{FRA} G. Franklin, private communication.
\bibitem{HAS} O. Hashimoto, private communication.
\end{references}
\end{document}